**Noncoding RNAs and deep learning neural network discriminate multi-cancer types**

Anyou Wang[1*], Rong Hai[1,2*], Paul J Rider[3], Qianchuan He[4]

[1]The Institute for Integrative Genome Biology, University of California at Riverside, Riverside, CA 92521, USA

[2]Department of Microbiology and Plant Pathology, University of California at Riverside, Riverside, CA 92521, USA

[3]Department of Pathobiological Sciences, School of Veterinary Medicine, Louisiana State University, Skip Bertman Drive, Baton Rouge, Louisiana, USA

[4]Public Health Sciences Division, Fred Hutchinson Cancer Research Center, Seattle, WA, 98109, USA

*Correspondence:

A.W.  anyou.wang@alumni.ucr.edu

R.H. ronghai@ucr.edu

**Running title**

Noncoding RNAs and AI classify multi-cancer types

**Abstract:**

**Background:** Detecting cancers at early stages can dramatically reduce mortality rates. Therefore, practical cancer screening at the population level is needed.

**Objective:** We develop a comprehensive detection system to classify all common cancer types

**Methods:** Integrate artificial intelligence deep learning neural network and noncoding RNA biomarkers selected from massive data.

**Results:** Our system can accurately detect cancer vs healthy object with 96.3% of AUC of ROC (Area Under Curve of a Receiver Operating Characteristic curve). Intriguinely, with no more than 6 biomarkers, our approach can easily discriminate any individual cancer type vs normal with 99% to 100% AUC. Furthermore, a comprehensive marker panel can simultaneously multi-classify all common cancers with a stable 78% of accuracy at heterological cancerous tissues and conditions.

**Conclusions:** This detection system provides a valuable practical framework for automatic cancer screening at population level.

**Key points:**

1) we develop a practical cancer screening system, which is a simple, accurate, affordable, and easy to operate

2) Our system binarily classify all cancers vs normal with >96% AUC

3) Total 26 Individual cancer types can be easily detected by our system with 99% to 100% AUC.

# 1 Introduction

Applications of modern scientific advances to cancer therapy have dramatically expanded cancer patient life expectancy[1–4]. One of the most successful practices is to detect cancers early and to remove them[5–10], which requires a practical screening system, a simple, accurate, affordable, and easy to operate system.

Numerous approaches have been recently proposed for cancer screening. Three remarkable innovations are promising. The first one is circulating tumor DNA (ctDNA) detection that measures bloodstream DNAs released from the dead tumor cells, but the amount of ctDNA is too low to being measured at early-stage tumors[7,9]. The second is a system consisting of two panels, a protein-based marker panel plus another mutation panel (8). Mutations are highly variable in humans and proteins are not good markers for cancers as we recently reported[11,12]. This resulted in wide range of variable accuracy in this system. The third is based on methylation[13]. Methylation is too expensive to measure and the methylation specificity for all types of cancers remains to be determined. Therefore, all these current proposals face challenges when applied to the real field. A practical screening system remains to be developed.

Here we developed a simple accurate framework to detect all common cancer types measured by TCGA (The Cancer Genome Atlas). Our system consists of artificial intelligence deep learning neural network and noncoding RNA biomarkers that are universal for all cancer types[11,12] and are easy to measure by cheap PCR.

# 2 Materials and methods

*2.1 General computational environment*

All data download, processing, computations and graphing were performed under Linux by using python 3.8 and R 3.6. TensorFlow 2.4.0 and Scikit-learn 0.24.0 were used for feature selection and deep learning neural network.

*2.2 Data resources*

All data were downloaded from TCGA as previously described[11]. Briefly, total 11,574 cancer samples for 36 cancer types were directly downloaded from TCGA publicly available data portal website. After filtering out cancer types with low sample size (samples <100), we kept a total of 26 cancer types with 9457 samples, including 8825 cancer samples and 632 normal samples for this study.

*2.3 Feature selection*

The 56 biomarkers for binarily classifying cancer vs normal shown in figure 1 were selected by our former study[11], in which ISURVIVAL model 2 that integrates stability-selection and first 200 principal components into Cox proportional hazards model to select unbiased universal cancer biomarkers.

Biomarkers for individual cancer types in figure 2 were collected by our previous study[11].

The comprehensive 325 marker panel shown in figure 3 were selected by inserting stability-selection into support vector machine implemented in feature selection from Scikit-learn. This frequency score >0.6 as described in our software FINET[14] was used as a cutoff.

*2.4 Data preparation*

Three independent data sets were prepared in this study, including test, validation, and training. These three groups were randomly split from total 9457 samples. The test set takes 20% (1891) of total 9457 samples for independently measuring final accuracy and AUC. Another 20% (1513) of the remaining samples (7566) was set for validation, and the remaining 6053 for training.

*2.5 Machine learning*

Deep learning neural network implemented in Keras Sequential library under Tensorflow was used throughout the whole study to estimate model accuracy, loss, and final AUC or accuracy. Batch size and epochs were set to 20 and 30 for all machine learning.

To avoid over-fitted, we set dropout (0.1) for each model layer for all models in this study. Two hidden layers with 30 and 60 units respectively were set for binary discrimination of cancer vs normal in figure 1 and individual cancer types in figure 2. Seven hidden layers with 240 units for each layer were set for multi-cancer classifiers in figure 3. Activation was set to relu for hidden layer. Adam was used for model optimizer.

Programming code and complete data are available on our project website ((https://combai.org/ai/cancerdetection/).

*2.6 Final graphing*

Final summary AUCs were drawn by using ggplot2 in R.

**3 Results**

3.1 **Noncoding RNAs work as universal cancer biomarkers to discriminate cancers from healthy objects**

One primary need of cancer screening is to discriminate cancers from healthy objects, regardless of cancer types. This requires a set of universal biomarkers for all types of cancers at all stages and conditions, which ensures cancer discrimination is not confounded by inappropriate specific variables. Our previous study developed algorithms to identify 56 noncoding RNAs universal for all cancer types after removing all specific effects such as cancer stage, age, sex, alcohol, smoking, and site location[11]. All biomarkers used in this study and detailed project info were deposited in our project website (https://combai.org/ai/cancerdetection/).

Here, we used these 56 noncoding RNAs as biomarkers and employed deep learning neural network (NN) from Keras Sequential library under TensorFlow v2.4.1 with 2 hidden layers (materials and methods) to binarily classify cancer vs normal. All cancer and normal samples measured by TCGA RNAseq project were used, 9457 total samples, including 8,825 cancer and 632 normal. To avoid over-fitted, we designed test and validation sets independent from the training samples and randomly split all 9457 samples into three sub-groups, test, validation, and training. The test set takes 20% (1891) of total 9457 samples for independently measuring AUC. The another 20% (1513) of rest (7566) was set for validation, and the remaining 6053 for training(materials and methods).

We examined the model accuracy, loss, and AUC for a series of biomarker numbers accumulated from 1 to 56. When the biomarker number accumulates to 13, the accuracy of training and validation reached 0.95 (Figure 1A left) respectively, and the loss declined to 0.14 and 0.15 (Figure 1A middle), and the AUC achieved 0.934 (Figure 1A right). When combined 51 biomarkers, the accuracy for both training and validation reached 0.96(Figure 1B left), and loss for training and validation went down to 0.10 and

0.15 respectively (Figure 1B middle), the AUC stabilized at 0.963 (Figure 1B right). All plots of this study results were shown in our website (https://combai.org/ai/cancerdetection/). Plotting AUC against the number of biomarkers provided a clear picture of the discrimination accuracy of our system(Figure 1C). While AUC was 0.75 for 1 biomarker, it first stabilized at 0.934 for 13 biomarkers, and elevated over 0.96 for >51 biomarkers. This indicated that our system can discriminate normal vs cancer with >0.96 AUC with less than 51 noncoding RNA markers.

**3.2 Individual cancer type discrimination**

Once a cancer sample is classified from normal as screened above, the next question is to determine its specific cancer type. By using the top 26 most common cancer types measured by TCGA, we previously employed elastic-net with stability-selection to select a set of noncoding RNA biomarkers to discriminate individual cancer types[11] but lacked a discrimination system optimizing AUC. With as many as 20 biomarkers, our previous elastic-net produced only 0.96 AUC. Here, we used deep learning neural network with this set of noncoding RNA biomarkers and built an accurate discrimination system (materials and methods). With only 1 biomarker, NN produced the accuracy for training and validation at 1.0 and 0.95 respectively for OV vs normal (Figure 2A, left), in which the loss for training and validation was closed to 0 and 0.1 respectively (Figure 2A middle) and AUC reached 100% for test data set for OV(Figure 2A right). The worst cases occurred for BRCA, which required 6 biomarkers to stabilize the accuracy and loss of both training and validation > 0.95 (Figure 2B left) and < 0.2 (Figure 2B middle) respectively, and 99.1% AUC (Figure 2B right) for test data. SARC also required 6 markers to achieve 99% AUC and it had only <80% AUC for 1 biomarker (Figure 2C). Within 6 biomarkers, all individual cancer types can be discriminated with 99% AUC (Figure 2C). 1 or 2 biomarkers were good enough (AUC from 99% to 100%) for most cancer types(Figure 2C). This suggested that noncoding RNAs plus NN can precisely classify any individual cancer types.

## 3.3 A comprehensive biomarker panel for multiple cancer classifiers

The subsequent challenge in cancer screening is to simultaneously detect all specific cancer types. This requires a comprehensive biomarker panel and a practical math model for multiple classifiers. Here we inserted stability-selection into support vector machine implemented in feature selection from Scikit-learn(0.24) and selected a panel of 325 noncoding RNA biomarkers (frequency score>0.6). This panel was designed to discriminate any type of cancers and normal, total 27 class units including 26 cancer types and 1 normal(materials and methods). Based on this panel, we built a NN model with 7 hidden layers (materials and methods). When 50 accumulated biomarkers were applied, the accuracy and loss for both training and validation reached >0.6 and close to 1 respectively (Figure 3A, upper and bottom panel). When biomarker number accumulated to 100, the accuracy and loss for both training and validation achieved >0.7 and loss declined to ~1(Figure 3B). The accuracy and loss for training and validation were stabilized at 0.78 and <1.0 for 171 accumulated biomarkers(Figure 3C). The accuracy curve for test data turned point at 100 biomarkers and stabilized at 0.78 after 171 biomarkers (Figure 3D). This suggested 171 biomarkers could be used to simultaneously detect all cancer types and normal.

## 4 Discussion

Our present study developed a practical system to detect all types of cancers by using noncoding RNAs and deep learning neural network. This system solved several practical problems existing in current cancer detection innovations, including imprecision, inconstancy, cost and immeasurability. The simpleness and high accuracy of our system result from deep learning neural network application. Compared with our previous results of 0.96 AUC with 20 biomarkers, the present system significantly increased AUC to 100% with only 1 or 2 biomarkers for most cancer types, and the worst case only needed 6 biomarkers. This significantly elevated the level of discrimination accuracy. In screening cancer vs normal, our system achieved 96% AUC with only 51 biomarkers. Noncoding RNAs work as robust markers for all heterogeneity cancers after removing confounding conditions[11,12], which ensures our system is constant and measurable. Noncoding RNAs are also easily measured by cheap, rapid, and sensitive PCR, and artificial intelligence neural network can be pre-programmed and trained, which enables any operators with no computer science background to operate our system. Therefore, our framework has great potential for population-based cancer screening.

Our comprehensive panel for multiple classifiers only reached 78%. This is expected because all samples come from heterological tissues and heterogeneity cancers. Once a unified material is applied, our framework would dramatically increase its accuracy. Overall, our panel provide a good idea to develop a multiple cancer classifier.

**Data and code availability**

All biomarker data and detailed project info was deposited (https://combai.org/ai/cancerdetection/)

**Acknowledgment**

This project was supported by University of California, Riverside initial funding

**Author contributions**

CONCEPTION: AW

INTERPRETATION OR ANALYSIS OF DATA: AW, RH, PR, HD, QH

PREPARATION OF THE MANUSCRIPT:AW

REVISION FOR IMPORTANT INTELLECTUAL CONTENT:AW

SUPERVISION:AW,RH

**Interest conflict**

No

**Figure legends**

**Figure 1. Binary discrimination of cancer and normal.** Noncoding RNA biomarkers and deep learning neural network accuratly discriminate all cancers from healthy objects. A, neural network model accuracy, loss and ROC curve of 13 accumulated biomarkers. B, neural network model accuracy, loss, and ROC curve of 51 accumulated biomarkers. C, AUC vs the number of accumulated biomarkers from 1 to 56.

**Figure 2. Binary classification of individual cancer types and normal.** A, only 1 biomarker can discriminate OV from normal with 100% AUC. B, Total 6 biomarkers were need for discriminating BRCA from normal with 99.1% AUC. C, AUC summary of all 26 individual cancer type discrimination.

**Figure 3. Multiple classifications of total 26 cancer types and normal.** A-C, model performance of 50, 100, and 171 accumulated biomarkers respectively. D, overall accuracy of multi-cancer classification with accumulated biomarkers.

Figure 1

A
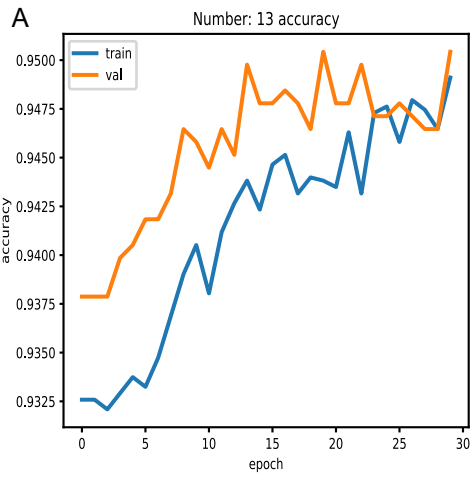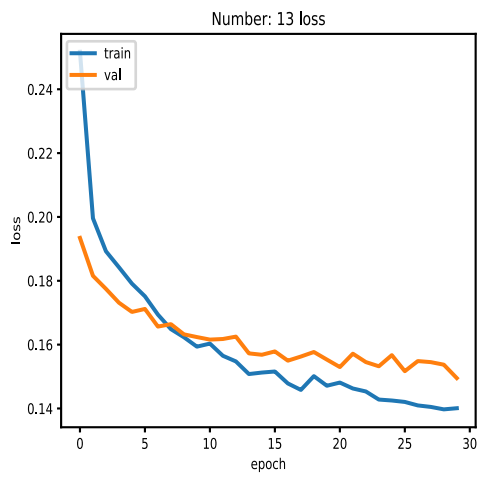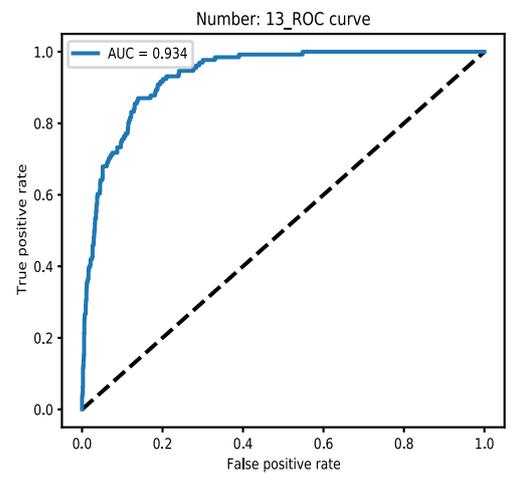

B
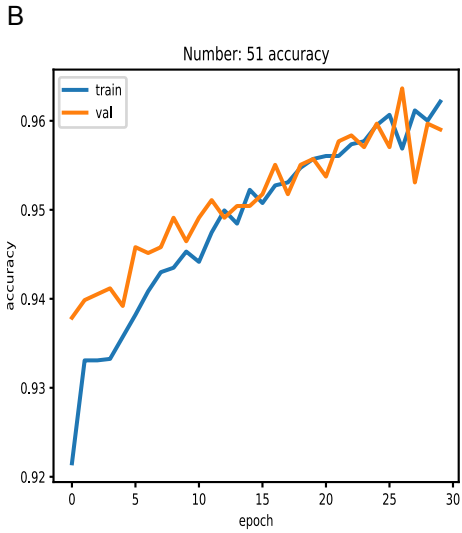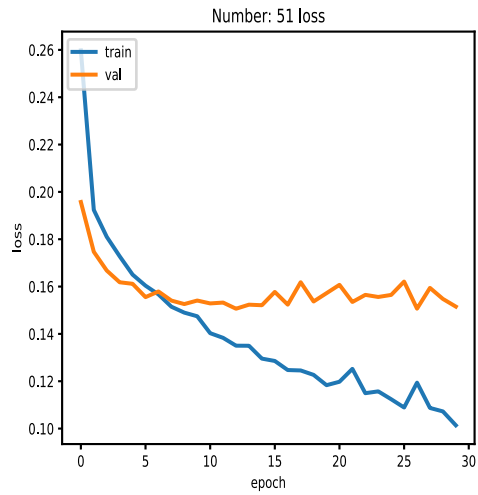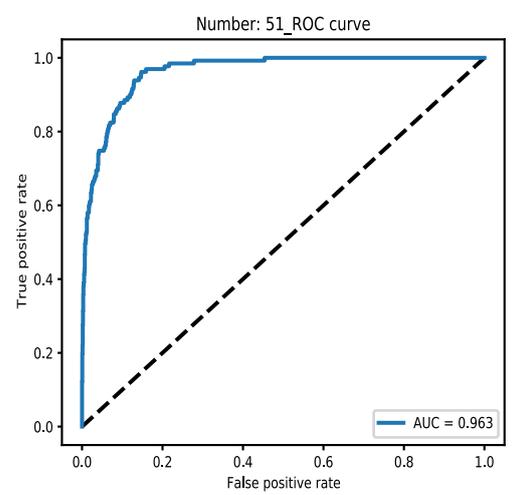

C
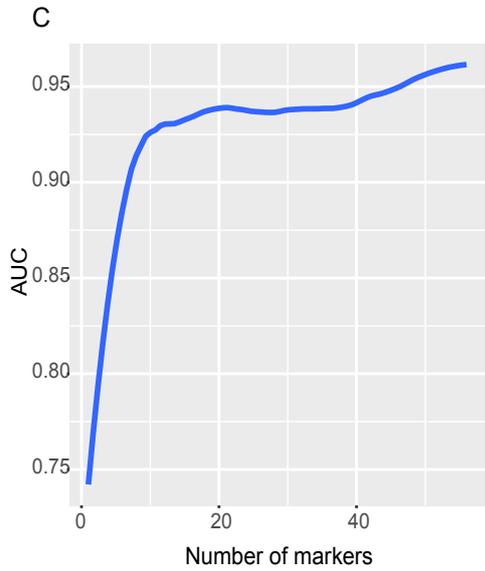

Figure 2

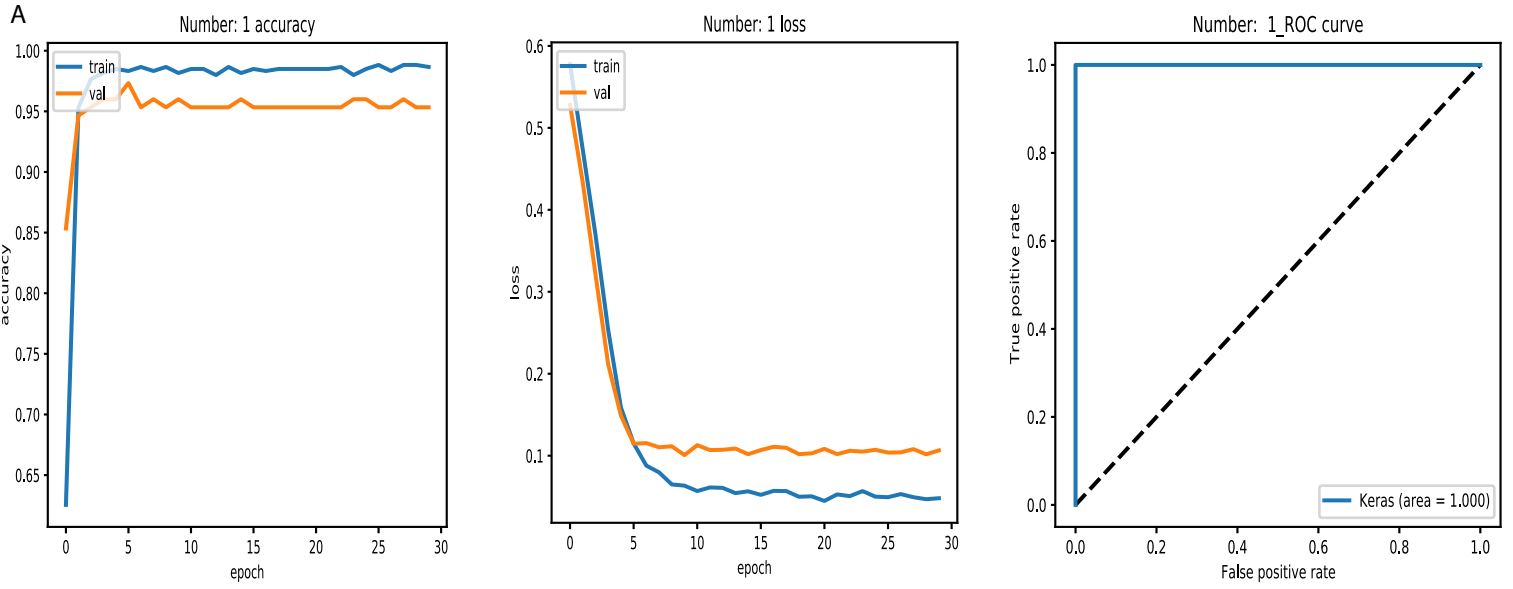

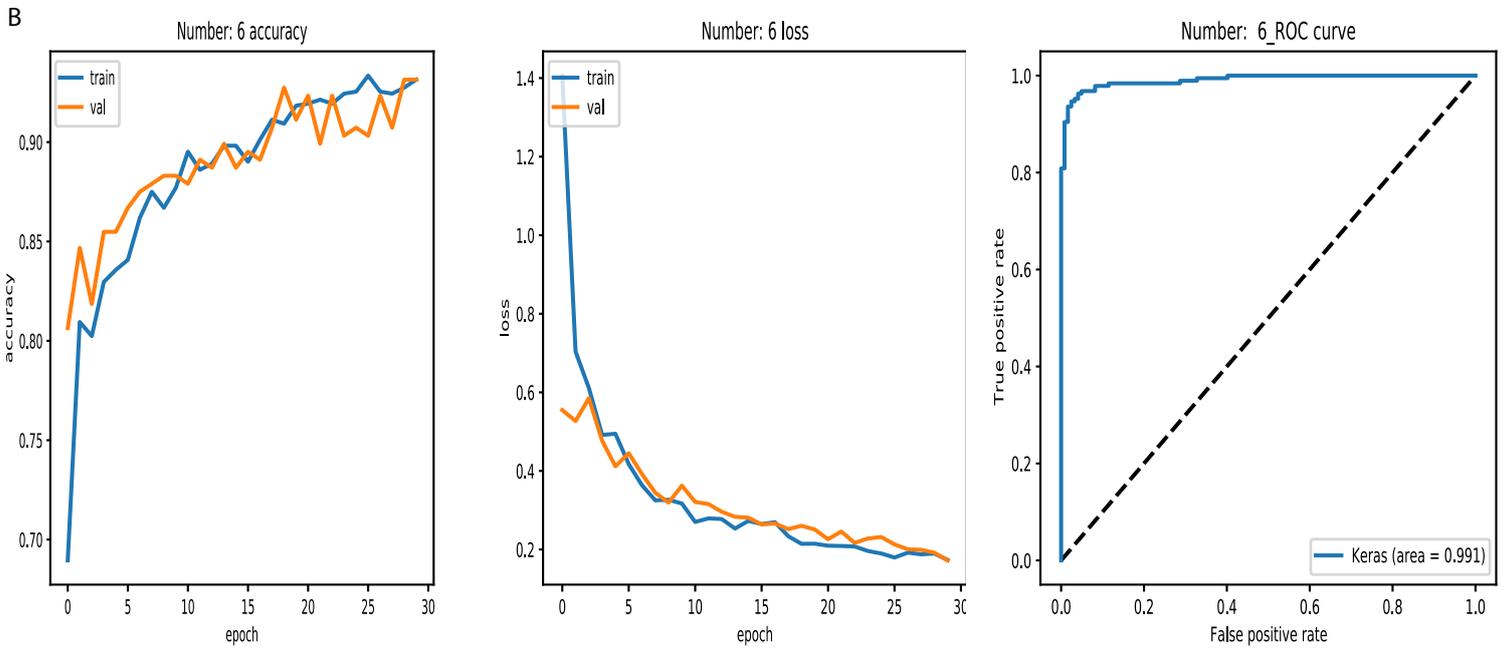

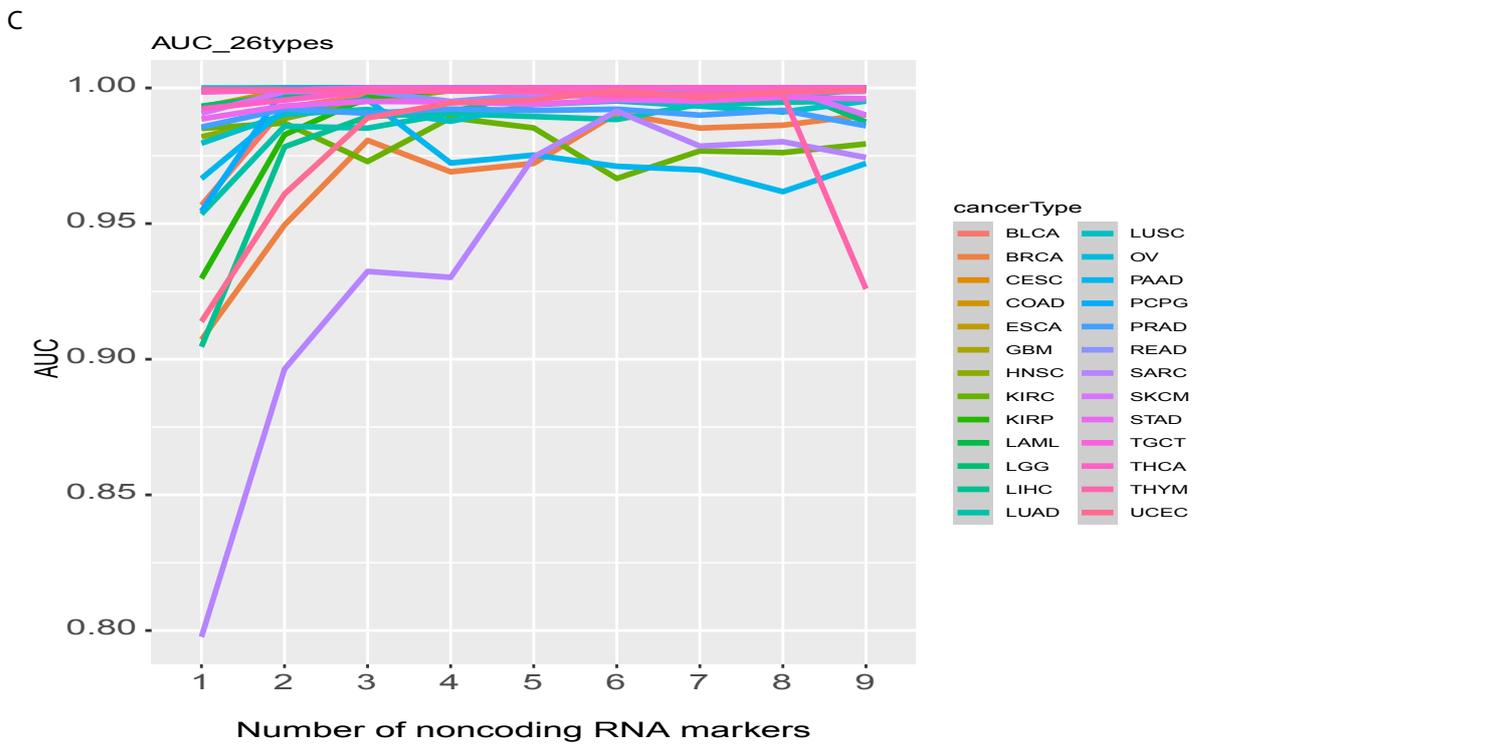

Figure 3

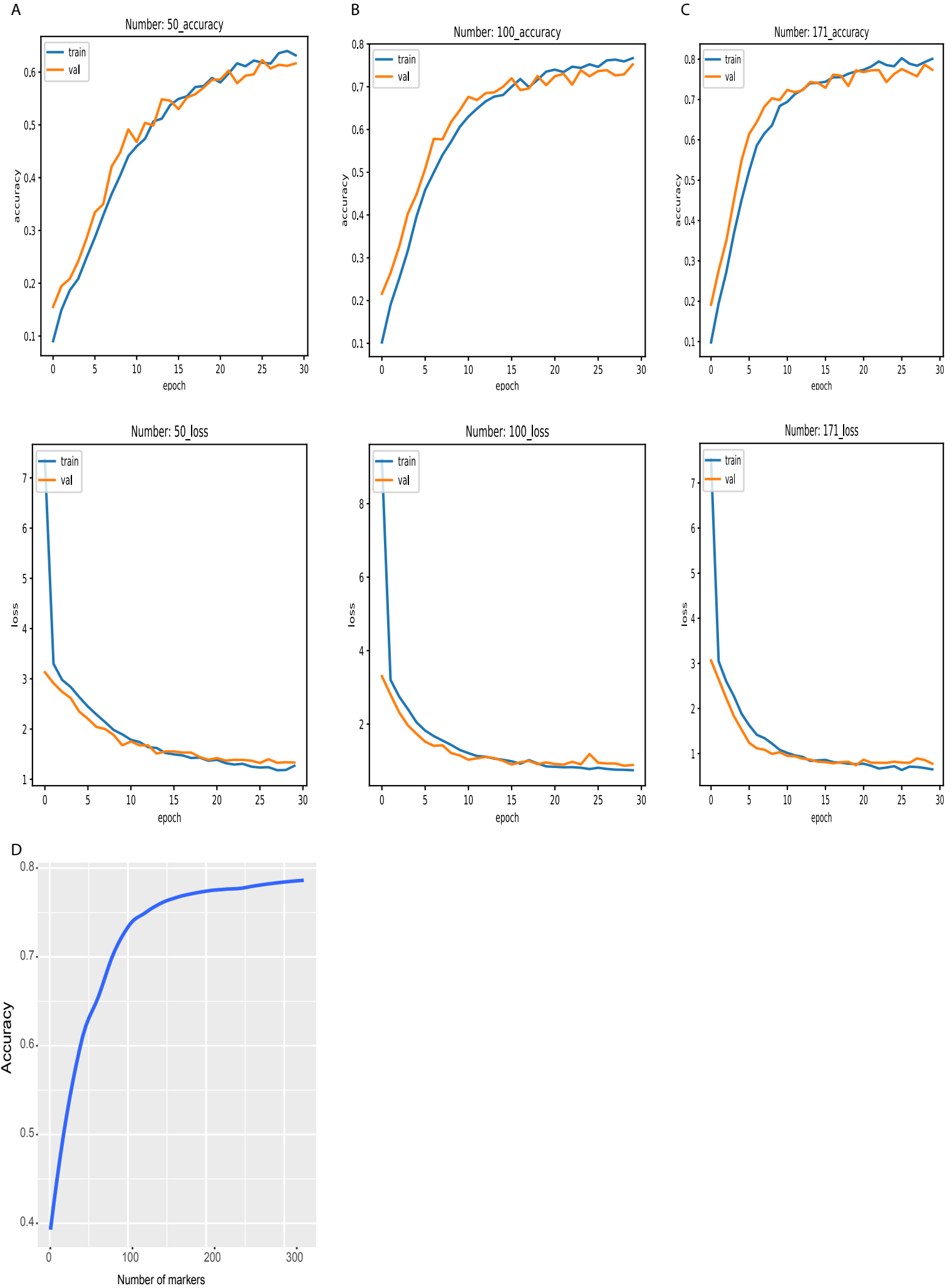